\documentclass[10pt,conference]{IEEEtran}

\usepackage{placeins}
\usepackage{afterpage}
\usepackage{mdframed}
\usepackage{xspace}
\usepackage{graphicx}
\usepackage{epstopdf}
\usepackage{color}
\usepackage{url}
\usepackage{subfig}
\usepackage{multirow}
\usepackage{array}
\usepackage{nowidow}
\usepackage{amssymb}
\usepackage{amsmath}
\usepackage{upquote}
\usepackage{flushend}
\usepackage{fancyvrb}
\usepackage{siunitx}
\usepackage{shortcuts}
\usepackage{pifont}
\usepackage{array}
\usepackage{mathtools}
\usepackage{multirow}
\usepackage{amstext}

\makeatother

\providecommand{\tabularnewline}{\\}
\usepackage{subfig}
\makeatother

\usepackage{listings, xcolor}

\definecolor{verylightgray}{rgb}{.97,.97,.97}

\lstdefinelanguage{Solidity}{
	keywords=[1]{anonymous, assembly, assert, balance, break, call, callcode, case, catch, class, constant, continue, contract, debugger, default, delegatecall, delete, do, else, event, export, external, false, finally, for, function, gas, if, implements, import, in, indexed, instanceof, interface, internal, is, length, library, log0, log1, log2, log3, log4, memory, modifier, new, payable, pragma, private, protected, public, pure, push, require, return, returns, revert, selfdestruct, send, storage, struct, suicide, super, switch, then, this, throw, transfer, true, try, typeof, using, value, view, while, with, addmod, ecrecover, keccak256, mulmod, ripemd160, sha256, sha3}, % generic keywords including crypto operations
	keywordstyle=[1]\color{blue}\bfseries,
	keywords=[2]{address, bool, byte, bytes, bytes1, bytes2, bytes3, bytes4, bytes5, bytes6, bytes7, bytes8, bytes9, bytes10, bytes11, bytes12, bytes13, bytes14, bytes15, bytes16, bytes17, bytes18, bytes19, bytes20, bytes21, bytes22, bytes23, bytes24, bytes25, bytes26, bytes27, bytes28, bytes29, bytes30, bytes31, bytes32, enum, int, int8, int16, int24, int32, int40, int48, int56, int64, int72, int80, int88, int96, int104, int112, int120, int128, int136, int144, int152, int160, int168, int176, int184, int192, int200, int208, int216, int224, int232, int240, int248, int256, mapping, string, uint, uint8, uint16, uint24, uint32, uint40, uint48, uint56, uint64, uint72, uint80, uint88, uint96, uint104, uint112, uint120, uint128, uint136, uint144, uint152, uint160, uint168, uint176, uint184, uint192, uint200, uint208, uint216, uint224, uint232, uint240, uint248, uint256, var, void, ether, finney, szabo, wei, days, hours, minutes, seconds, weeks, years},	% types; money and time units
	keywordstyle=[2]\color{teal}\bfseries,
	keywords=[3]{block, blockhash, coinbase, difficulty, gaslimit, number, timestamp, msg, data, gas, sender, sig, value, now, tx, gasprice, origin},	% environment variables
	keywordstyle=[3]\color{violet}\bfseries,
	identifierstyle=\color{black},
	sensitive=false,
	comment=[l]{//},
	morecomment=[s]{/*}{*/},
	commentstyle=\color{gray}\ttfamily,
	stringstyle=\color{red}\ttfamily,
	morestring=[b]',
	morestring=[b]"
}

\begin{document}
\title{Slither: A Static Analysis Framework For Smart Contracts}
%
%\titlerunning{Abbreviated paper title}
% If the paper title is too long for the running head, you can set
% an abbreviated paper title here
%
%\author{Anonymous %\inst{1}\orcidID{0000-1111-2222-3333} \and
%Second Author\inst{2,3}\orcidID{1111-2222-3333-4444} \and
%Third Author\inst{3}\orcidID{2222--3333-4444-5555}}
%
%\authorrunning{Anonymous et al.}
% First names are abbreviated in the running head.
% If there are more than two authors, 'et al.' is used.
%
%\institute{Princeton University, Princeton NJ 08544, USA \and
%Springer Heidelberg, Tiergartenstr. 17, 69121 Heidelberg, Germany
%\email{lncs@springer.com}\\
%\url{http://www.springer.com/gp/computer-science/lncs} \and
%ABC Institute, Rupert-Karls-University Heidelberg, Heidelberg, Germany\\
%\email{\{abc,lncs\}@uni-heidelberg.de}
%}

\author{\IEEEauthorblockN{Josselin Feist}
\IEEEauthorblockA{\textit{Trail of Bits}\\
New York, New York\\
josselin@trailofbits.com
}
\and
\IEEEauthorblockN{Gustavo Grieco}
\IEEEauthorblockA{\textit{Trail of Bits}\\
New York, New York\\
gustavo.grieco@trailofbits.com
}
\and
\IEEEauthorblockN{Alex Groce}
\IEEEauthorblockA{\textit{Northern Arizona University} \\
Flagstaff, Arizona\\
agroce@gmail.com}
}
\maketitle              % typeset the header of the contribution

\begin{abstract}

This paper describes Slither, a static analysis framework designed to provide rich information about Ethereum smart contracts.
It works by converting Solidity smart contracts into an intermediate representation called SlithIR.
SlithIR uses Static Single Assignment (SSA) form and a reduced instruction set to ease implementation of analyses while preserving semantic information that would be lost in transforming Solidity to bytecode. 
Slither allows for the application of commonly used program analysis techniques like dataflow and taint tracking. 
Our framework has four main use cases: 
(1) automated detection of vulnerabilities,
(2) automated detection of code optimization opportunities,
(3) improvement of the user's understanding of the contracts, and
(4) assistance with code review.  

In this paper, we present an overview of Slither, detail the design of its intermediate representation, and evaluate its capabilities on real-world contracts. 
We show that Slither's bug detection is fast, accurate, and outperforms other static analysis tools at finding issues in Ethereum smart contracts in terms of speed, robustness, and balance of detection and false positives. We compared tools using a large dataset of smart contracts and manually reviewed results for $1000$ of the most used contracts.  
\end{abstract}

\section{Introduction}\label{sec:intro}

%Smart contracts are everywhere!
A growing number of industries are using blockchain platforms to perform trustless computation using smart contracts. 
Applications ranging from financial services to supply chains, and from logistics to healthcare are being developed to rely on blockchain technologies. 
One of the most popular underlying technologies is the Ethereum smart contact. 
These contracts are typically written in a high-level programming language called Solidity, then compiled to Ethereum  Virtual Machine (EVM) assembly instructions for blockchain deployment. 
Often, the deployed smart contract's code is insecure: software vulnerabilities are regularly identified, and have been exploited by malicious actors, resulting in millions of dollars in damages and harm to the reputation of blockchain systems.

%We can find the vulnerabilities using static analysis tools

In the last few years, a variety of tools and frameworks to analyze and find vulnerabilities in Ethereum smart contracts were developed based on static and dynamic analysis. 
These tools are based on popular program testing techniques such as fuzzing, symbolic execution, taint tracking, and static analysis. 

Static analysis is one of the most effective ways to detect potential issues in contracts. 
Typically, static analysis tools work by analyzing the Solidity source code or a disassembled version of the compiled contract. Then, they transform the contract code into an internal representation, more suitable for the analysis and detection of common security issues.  
%Because these tools can be used through the smart contract development lifecycle to detect vulnerabilities it is imperative that they be as precise and effective as possible.

%Static framework have problems
%However, static analysis tools on Ethereum are far from perfect. 
While modern compilers, such as clang, offer various APIs on top of which third-party analyzers can be built, the Solidity compiler fails to offer the same features. 
Ideally, a static analysis framework for Ethereum smart contracts should have the following properties:
%Thus, other static analysis tools to check their code in third-party software. Ideally, a static analysis framework for Ethereum smart contract should have the following properties:

\begin{enumerate}
	\item \textbf{Correct level of abstraction}: if the framework is too abstract, it can be hard to introduce accurate semantics that capture common usage patterns. 
    Conversely, if the framework is too narrowly focused on the detection of certain issues only, it can be difficult to add new detectors or analyses.  
	\item \textbf{Robustness}: it should parse and analyze real-world code without crashing.
    %It is very important to ensure that the analysis performed in a smart contract is complete: all functions should be scanned properly to find issues in each line of code. A silent error in a smart contract static analyzer could hide a potential security or correctness issue with catastrophic consequences if the contract is deployed and cannot be upgraded easily.
	\item \textbf{Performance}: analysis should be fast, even for large contracts, so as to integrate easily into development tools like IDEs, or into continuous integration checks that run at every commit. 
%	\item \textbf{Performance}: it should be fast, even for large contracts. This is important for two main reasons: (a) a quick analysis is easier to integrate into development tools like an IDE to highlight errors while the developer is modifying the contract, or into continuous integration checks that should run at every commit. And (b) if the static tool takes a long time to run, it could be more effective to run a dynamic analysis tool, since they employ a more precise approach.
	\item \textbf{Accuracy}: it should allow for the development of detectors that find most potential issues while maintaining a low false positive rate. If the number of false positives is very high, it can require a manual audit of the entire contract, defeating its utility.
	\item \textbf{Batteries included}: it should include a set of common analyses and issue detectors that are useful for most contracts. This will appeal both to security engineers looking to extend the framework and to code auditors looking for issues to report.

\end{enumerate}

This paper introduces Slither, an open-source static analysis framework. Our tool provides rich information about Ethereum smart contracts and has the  critical properties enumerated above. 
%Slither relies on the Solidity compiler to parse the smart contract and employs a multi-step procedure for analysis. 
%First, it coverts the smart contract into an intermediate representation called SlithIR.
Slither uses its own intermediate representation, SlithIR, designed to make static analyses on Solidity code straightforward. 
%This representation uses Static Single Assessment (SSA), and its reduced set of instructions makes static analysis directly on Solidity code straightforward.  
It applies widely used program analysis techniques such as dataflow and taint tracking to extract and refine information. % from every line of code. 
%The framework includes a set bugs and optimizations detectors, as well as several outputs to assist the user code understanding.
%Finally, it passes the information collected to specific modules designed to detect common issues. 
%Finally, it passes the information collected to specific modules designed to detect common issues. 
While Slither is built as a security-oriented static analysis framework, it is also used  to enhance the user's understanding of smart contracts, assist in code reviews, and detect missing optimizations.

We summarize our contribution as follows:
\begin{itemize}
\item We present Slither, a framework for static analysis of Solidity contracts.
\item We detail the design of Slither's intermediate representation and analyzers.
\item We evaluate and compare the performance, robustness, and accuracy of Slither on a large set of actively used smart contracts. 
%\item We open source both Slither and our dataset so that others may validate and improve upon our results.
\end{itemize}

Both Slither and our dataset are open source, so that others can validate and improve on our results.  The rest of the paper is organized as follows. Section~\ref{sec:related-work} presents related work.
Section~\ref{sec:overview} introduces a high-level overview of Slither's architecture, and Section~\ref{sec:slithir} details the intermediate representation language SlithIR. 
In Section~\ref{sec:evaluation}, we explain how we performed an evaluation and comparison of Slither against state-of-the-art tools and discuss the results. 
Finally, Section~\ref{sec:conclusions} summarizes the paper and discusses future work.

\section{Related Work}\label{sec:related-work}

Several frameworks have been created to allow users and security researchers to analyze Ethereum smart contracts and detect potential issues. 
They are based either on static or dynamic analysis.

\subsection{Static Analysis}
These tools rely on the analysis of the code without executing it to detect issues in the smart contract.
\emph{Securify} is one of these tools~\cite{securify}, developed by the SRI Systems Lab (ETH Zurich). 
It works at the bytecode level: first, it parses and decompiles the EVM bytecode, then it translates the resulting code to \emph{semantic facts} using static analysis. 
Finally, it matches the facts with a list of predefined patterns to detect common issues. 
\emph{Securify} is open-source. 
It was implemented using Java and stratified Datalog. 
\emph{SmartCheck} is another static analysis tool~\cite{smartcheck}, developed by SmartDec. 
It works by translating directly from the Solidity source code into an XML-based intermediate representation. 
It then checks the intermediate representation against \emph{XPath} patterns to identify potential security, functional, operational, and development issues. 
It is also implemented using Java. 
\emph{Solhint} is a tool for linting Solidity code~\cite{solhint-code}, developed by ProtoFire. 
It aims to provide both security and style guide validations. 
It is open-source and it is implemented in NodeJS using the \emph{SolidityJ} parser~\cite{solidityj-code}. 
Other notable static analysis frameworks include Vandal~\cite{Brent2018VandalAS} and EtherTrust~\cite{ethertrust}.

GASPER~\cite{chen2017under} and GasReduce~\cite{chen2018under} use static analysis to detect potential optimizations in contracts, GASPER at a high-level (e.g. dead code) and GasReduce at the bytecode instruction pattern level.  Both focus on dead code and loop optimizations, rather than, e.g., the could-be-declared-constant optimization we discuss below, so these approaches are largely orthogonal to Slither's optimization patterns.

\subsection{Dynamic Analysis}
These tools rely on execution of the contract, leveraging symbolic execution, taint tracking, and fuzzing to discover vulnerabilities. 
Oyente was one of the first~\cite{oyente} tools for analysis and detection of security issues in Ethereum smart contracts. 
It is developed by Melonport and its code is open-sourced~\cite{oyente-code}. 
Manticore is an open-source symbolic execution tool for analysis of Ethereum smart contracts and binaries created by Trail of Bits~\cite{manticore-code}. 
Echidna~\cite{echidna-code}, another product of Trail of Bits, is a property-based testing tool designed for fuzzing Ethereum smart contracts.
Mythril Classic is an open-source security analysis tool for Ethereum smart contracts created by ConsenSys~\cite{mythril-code}. It uses concolic analysis, taint analysis, and control flow checking to detect a variety of security vulnerabilities.
Finally, TeEther is an automatic exploit generation tool for certain types of vulnerabilities in Ethereum smart contracts created by Krupp and Rossow~\cite{teether}. 
Its source code is not available at the time of this writing, but its authors promised to make their tool open-source 90 days after its presentation at Usenix 2018.

\section{Slither}\label{sec:overview}

Slither is a static analysis framework designed to provide granular information about smart contract code and the flexibility necessary to support many applications.
The framework is currently used for the following:

\begin{itemize}
	\item \textbf{Automated vulnerability detection}: a large variety of smart contract bugs can be detected without user intervention. %(Section \ref{subsection:vulnerabilities_detection}) 
	\item \textbf{Automated optimization detection}: Slither detects code optimizations that the compiler misses \cite{chen2017under}. %(section \ref{subsection:optimizations_detection})
	\item \textbf{Code understanding}: printers summarize and display contracts' information to aid in the study of the codebase. %(section \ref{subsection:code_understanding})
	\item \textbf{Assisted code review}: through its API, a user can interact with Slither. %(section \ref{subsection:assisted_code_review})
\end{itemize}

\begin{figure*}[t]
\centering
\includegraphics[width=0.95\textwidth]{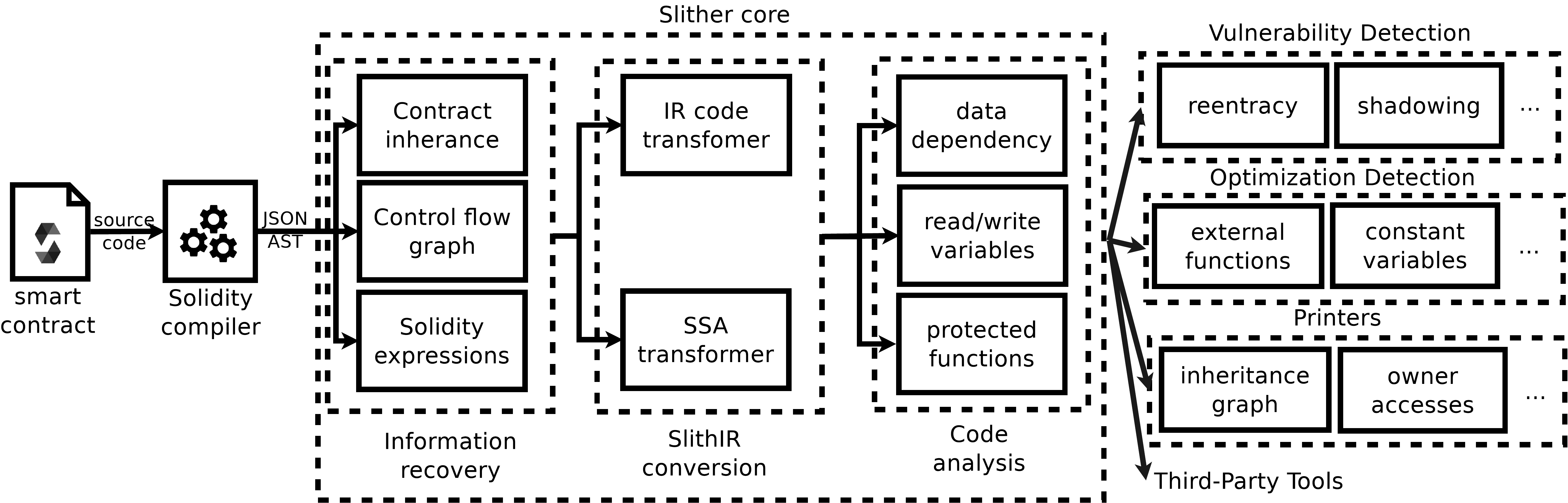}
\caption{Slither overview}
\label{fig:overview}
\end{figure*}

Slither analyzes contracts using static analysis in a multi-stage procedure.
Slither takes as initial input the Solidity Abstract Syntax Tree (AST) generated by the Solidity compiler from the contract source code.
In the first stage, Slither recovers important information such as the contract's inheritance graph, the control flow graph (CFG), and the list of expressions. 
Next, Slither transforms the entire code of the contract to SlithIR, its internal representation language.
SlithIR uses static single assessment (SSA) to facilitate the computation of a variety of code analyses.
During the third stage, actual code analysis, Slither computes a set of pre-defined analyses that provide enhanced information to the other modules. 
%analyses such as taint tracking or identification of protected functions that are essential to build precise and fast detectors.
Figure \ref{fig:overview} summarizes all these stages.

\subsection{Built-in Code Analyses}

\begin{itemize}

\item \textbf{Read/Write}

Slither identifies the reads and the writes of variables.
For each contract, function, or node of the control flow graph, it is possible to retrieve the variables read or written, and filter by the type of variables (local or state). 
For example, it is possible to know the state variables written from a specific function, or find which functions write to a given variable.
Several detectors are based on this information, such as those for uninitialized variables and reentrancy.

\item \textbf{Protected functions}
A recurrent pattern in smart contract design is the use of ownership to protect access to  functions. 
The thinking behind this pattern is that a specific user (or set of users), called the \textit{owner(s)}, can perform privileged operations. 
%Figure~\ref{fig:isowner} shows this pattern.
For instance, only an owner can mint new tokens.
Modeling the protection of functions lowers the number of false positives.
In order to detect unprotected functions, Slither checks for cases
where (1) the function is not the constructor and (2) the address of
the caller, \textit{msg.sender}, is not directly used in a
comparison. This heuristic can give false positives and false negatives, but our experience shows that it is effective in practice.

%\begin{figure}
%\begin{lstlisting}{language=Solidity}
%contract Owner{
% address owner = msg.sender;
%
% // Can be only executed by the contract's owner
% modifier isOwner(){
%   require(msg.sender == owner);
%   _;
% }
%}
%  
%\end{lstlisting}
%\caption{isOwner example}
%\label{fig:isowner}
%\end{figure}

\item \textbf{Data dependency analysis}
Slither computes the data dependencies of all variables as a pre-stage using the SSA representation.
The dependencies are first analyzed in the context of each function. 
Then a fixpoint is computed across all the functions of the contract,
to determine whether there is a dependency in a multi-transaction context.
Slither classifies some variables as \textit{tainted}, which means that the variable is dependent on a user-controlled variable, and, therefore, can be influenced by the user.
Finally, the data dependency takes into account the protected function heuristic. 
As a result, it is possible to select dependencies according to the user's privilege level.
This granular information about data dependency is essential to writing accurate vulnerability detectors.
\end{itemize}

\subsection{Automated Vulnerability Detection}
\label{subsection:vulnerabilities_detection}
The open source version of Slither contains more than twenty bug detectors, including:
\begin{itemize}
\item Shadowing: Solidity enables the shadowing of most of the contract's elements, such as local and state variables, functions, or events.  Shadowed elements may not refer to the objects expected by the contract author.
\item Uninitialized variables: Uninitialized state or local variables are a common source of errors in programming languages.
\item Reentrancy: The reentrancy vulnerability (see
  Section~\ref{subsec:background-reentrancy}) is well known.
\item A variety of other known security issues, such as suicidal contracts, locked ether, or arbitrary sending of ether \cite{maian}.
\end{itemize}

Closed-source detectors extend Slither's capabilities to detect more advanced bugs, such as race conditions, unprotected privileged functions, or incorrect token manipulations.
\subsection{Automated Optimization Detection}
\label{subsection:optimizations_detection}
Slither can detect code patterns that lead to costly code execution and deployment, a topic of considerable interest for smart contracts where there is a clear financial cost associated with inefficient code \cite{chen2017under}. 
The framework includes detectors for:
\begin{itemize}
\item Variables that should be declared as constants. Constant variables are optimized by the compiler\footnote{\url{https://solidity.readthedocs.io/en/v0.5.2/contracts.html#constant-state-variables}}
, consume less gas, and do not take space in storage.
\item Functions that should be declared as externals. External functions allow the compiler to optimize the code\footnote{\url{https://solidity.readthedocs.io/en/v0.5.2/contracts.html#visibility-and-getters}}.
\end{itemize}

\subsection{Code Understanding}
\label{subsection:code_understanding}

Slither includes \emph{printers}, which allow users to quickly understand what a contract does and how it is structured.
Open source printers include:
\begin{itemize}
\item The exportation of different graph-based representations, including the inheritance graph, the control flow graph, and the call graph of each contract.
\item A human-readable summary of the contracts, including the number of issues found and information about the code quality, such as cyclomatic complexity, or a minting restriction for ERC20 tokens.
\item A summary of the authorization accesses and the variables that can be changed by the contract's owner.
\end{itemize}

\subsection{Assisted Code Review}
\label{subsection:assisted_code_review}
Users can build third-party scripts and tools using Slither Python API\footnote{\url{https://github.com/crytic/slither/wiki/API-examples}}.
A custom script can target contract-specific needs.
For example, a user can ensure that a specific variable is never tainted by the parameter of a given function, or that a function is reachable only via legitimate entry points.
Third-party tools can use Slither internals to build more advanced analyses, such as symbolic execution on top of SlithIR, or a conversion from SlithIR to another intermediate representation, such as LLVM.

\subsection{Implementation}

Slither is implemented in $16$K lines of Python 3 code and is available on GitHub\footnote{\url{https://github.com/crytic/slither}}. 
It has support for continuous integration and developer toolboxes (such as Truffle
%\footnote{\url{https://truffleframework.com/}}
and Remix).
%\footnote{\url{http://remix.ethereum.org/}}).
It has minimal dependencies and only relies on a recent version of the Solidity compiler to parse the AST of the contract under analysis.

\section{SlithIR}\label{sec:slithir}

SlithIR is the hybrid intermediate representation used in Slither to represent Solidity code. 
Each node of the control flow graph can contain up to a single Solidity expression, which is converted to a set of SlithIR instructions.  This representation makes implementing analyses easier, without losing the critical semantic information contained in the Solidity source code.

%\textbf{Syntax} SlithIR defines variables to be either Lvalues (LV) o Rvalues (RV). The syntax for both types of variables is the following:
%
%\begin{align*}
%\text{LV} & \Coloneqq & \text{StateVar }|\text{ LocalVar} &  & \text{RV} & \Coloneqq & \text{StateVar }|\text{ LocalVar}\\
% & | & \text{ TupleVar }|\text{ TempVar} &  &  & | & \text{Const }|\text{ SolVar}\\
% & | & \text{RefVar} &  &  & | & \text{TempVar }|\text{ RefVar}
%\end{align*

\subsection{Instruction Set}
SlithIR uses fewer than 40 instructions. %, as shown in Tables \ref{subtable:slithir1} and \ref{subtable:slithir2}. 
It has no internal control flow representation and relies on Slither's control-flow
graph structure (SlithIR code is associated with each
node in the graph).
The following gives a high-level description of some notable instructions; the complete descriptions is available at~\cite{slither-slithir}.

\subsubsection{Notation} $LV$ and $RV$ respectively represent a variable that is assigned (left-value) and a variable that is read (right-value).
A variable can be a Solidity variable or a temporary variable created by the intermediate representation.

\subsubsection{Arithmetic Operations} are represented as either a Binary or Unary operator:
\begin{itemize}
\item \texttt{LV $=$ RV BINARY RV}
\item \texttt{LV $=$ UNARY RV}
\end{itemize}

\subsubsection{Mappings and Arrays:} Solidity allows the manipulation of mappings and arrays that are accessed through dereferencing.
SlithIR uses a specific variable type, called $REF$ (ReferenceVariable) to store the result of a dereferencing.
The index operator allows dereferencing of a variable:
\begin{itemize}
\item \texttt{REF $\leftarrow$ Variable [Index]}
\end{itemize}
\subsubsection{Structures:} Access to a structure is done through the member operator:
\begin{itemize}
\item \texttt{REF $\leftarrow$ Variable $\cdot$ Member} 
\end{itemize}
\subsubsection{Calls:} Slither gives in-depth information about calls and possesses nine call instructions:
\begin{itemize}
\item \texttt{LV $=$ L\_CALL Destination Function [ARG..]} (low-level Solidity call)
\item \texttt{LV $=$ H\_CALL Destination Function [ARG..]} (high-level Solidity call)
\item \texttt{LV $=$ LIB\_CALL Destination Function [ARG..]} (library call)
\item \texttt{LV $=$ S\_CALL Function [ARG..]} (call to a inbuilt-Solidity function)
\item \texttt{LV $=$ I\_CALL Function [ARG..]} (call to an internal function)
\item \texttt{LV $=$ DYN\_CALL Variable [ARG..]} (call to an internal dynamic function)
\item \texttt{LV $=$ E\_CALL Event [ARG..]} (event call)
\item \texttt{LV $=$ Send Destination} (Solidity $send$)
\item \texttt{Transfer Destination} (Solidity $transfer$)
\end{itemize}
Some calls can have additional arguments, for example, \texttt{H\_CALL}, \texttt{L\_CALL}, Send and Transfer can have $Value$ associated to the call, representing the amount of ether for the transaction.

\subsubsection{Additional Instructions} include \texttt{PUSH} for array manipulation, \texttt{CONVERT} for type conversion, and operators to manipulate tuples. 

%For example, all the operations that write to a variable belong to the OperationWithLValue group.
%MThe instructions hierarchy simplifies the build of analyses, for example, an analysis that requires to track only write operation can be applied only on instruction of OperationWithLvalue type.

\subsubsection{Example}
Figure~\ref{fig:slithir-example} is the SlithIR representation of the code showed in Figure~\ref{fig:slithir-code-example}.

\begin{figure}
	\begin{lstlisting}[language=solidity]
using SafeMath for uint;
mapping(address => uint) balances;

function transfer(address to, uint val) public{
    balances[msg.sender] = balances[msg.sender].min(val);
    balances[to] = balances[to].add(val);
}
	\end{lstlisting}
	\caption{Solidity code example}
	\label{fig:slithir-code-example}
\end{figure}

\begin{figure}
	\begin{lstlisting}[basicstyle=\scriptsize]
Function transfer(address,uint256)
	Solidity: balances[msg.sender] = balances[msg.sender].sub(val)
	SlithIR:
		REF_0(uint256) -> balances[msg.sender]
		REF_1(uint256) -> balances[msg.sender]
		TMP_1(uint256) = LIB_CALL SafeMath.sub(REF_1, val)
		REF_0 := TMP_1(uint256) // dereferencing
	Solidity: balances[to] = balances[to].add(val)
	SlithIR:
		REF_3(uint256) -> balances[to]
		REF_4(uint256) -> balances[to]
		TMP_3(uint256) = LIB_CALL, dest:SafeMath.add(REF_4, val)
		REF_3 := TMP_3(uint256) // dereferencing
	\end{lstlisting}
	\caption{SlithIR code for Figure~\ref{fig:slithir-code-example}}
	\label{fig:slithir-example}
\end{figure}

\subsection{SSA}
Static Single Assignment form~\cite{SSA} (SSA) is a representation commonly used in compilation and static analysis in general.
It requires that each variable is assigned only one time and defined prior to its usage; e.g., {\tt x = 3; y = x++; x = y + x} becomes {\tt x$_1$ = 3; y$_1$ = x$_1$; x$_2$ = x$_1$ + 1; x$_3$ = y$_1$ + x$_2$}.
One of the main advantages of SSA is to allow def-use chains to be easily computed, making data-dependency analyses straightforward.  Using
SSA form also enables more aggressive future analyses.  For example,
because gas limits force a bound on computational cost of contract
execution, bounded model checking~\cite{BMC} using SAT/SMT is a natural fit for contract analysis.  SlithIR's SSA representation plus graph-based control
flow is very similar to the representation used
in the successful CBMC~\cite{CBMCp} tool for C.

Slither stores two SlithIR versions: with and without SSA.

\subsubsection{State Variables}
A so-called $\phi$ function on a node indicates that a variable has
multiple potential definitions and is a key element of SSA form~\cite{SSA}.
One particularity of smart contracts is to heavily rely on state variables, which act as global variables.
At the beginning of a function, the value of a state variable can be its initial value, or the value after the execution of any function.
Additionally, an external call may re-enter, which can allow a state variable to change. 
As a result, a $\phi$ function is placed at the entry of each function, and after the external calls, for each state variable read by the function.

%\begin{figure}
%	\begin{lstlisting}
%contract C{
%  uint my_var;
%
%  function my_func(address callback) external returns(uint){
%      my_var = my_var + 1;
%      callback.call();
%      my_var = my_var + 1;
%  }
%
%  function inc() external{
%     my_var = 1;
%  }
%}
%	\end{lstlisting}
%	\caption{TODO}
%	\label{fig:ssa_state_variables_callback}
%\end{figure}
%
%\begin{figure}
%	\begin{lstlisting}[language=bash]
%
%Contract C
%	Function my_func(address)
%		Expression: my_var = my_var + 1
%		IRs:
%			TMP_0(uint256) = my_var_0 + 1
%			my_var_1(uint256) := TMP_0(uint256)
%		Expression: callback.call()
%		IRs:
%			TMP_2(bool) = LOW_LEVEL_CALL, dest:callback_0, function:call, arguments:[]  
%			my_var_2(uint256) := ϕ(['my_var_4', 'my_var_1', 'my_var_3'])
%		Expression: my_var = my_var + 1
%		IRs:
%			TMP_3(uint256) = my_var_2 + 1
%			my_var_3(uint256) := TMP_3(uint256)
%	Function inc()
%		Expression: my_var = 1
%		IRs:
%			my_var_4(uint256) := 1(uint256)
%
%	\end{lstlisting}
%	\caption{TODO (improve me)}
%	\label{fig:ssa_state_variables_callback_output}
%\end{figure}

%\begin{itemize}
%	\item Beginning of the functions
%	\item After each external call
%\end{itemize}

\subsubsection{Alias Analysis}
Solidity allows local variables to refer to state variables, as shown in Figure~\ref{fig:ssa_reference}.
Such variables are called \textit{storage references}.
A write operation to a local variable can, therefore, refer to a write to multiple state variables.
Slither computes an alias analysis to find all the possible targets of the \textit{storage reference} and gives the information to the SSA engine.
As a result, $\phi$ functions can be correctly placed after the write to a storage reference.
In the example in Figure~\ref{fig:ssa_reference}, two $\phi$ functions will be placed at the expression $ref.val += 1;$ to indicate that $a$ and $b$ are updated.

\begin{figure}
	\begin{lstlisting}[language=solidity]
struct MyStructure {
    uint val;
}

MyStructure private a;
MyStructure private b;

function increase(bool useb) external {
    MyStructure storage ref = a;
    if(useb){
        ref = b;
    }
    ref.val += 1;
}
	\end{lstlisting}
	\caption{Smart contract using storage references}
	\label{fig:ssa_reference}
\end{figure}

%\subsection{Example}
%
%\begin{tabular}{|l|l|}
%\hline 
%Solidity & SlithIR\tabularnewline
%\hline 
%contract UsafeMath \{ & \tabularnewline
%~~~~function add(uint256 a, uint256 b) \{ & \tabularnewline
%\multirow{2}{*}{~~~~~~~~return a + b;} & TMP\_0(uint256) = a + b\tabularnewline
% & RETURN TMP\_0\tabularnewline
%~~~~\} & \tabularnewline
%\} & \tabularnewline
%\hline 
%\end{tabular}

\subsection{Comparison to Other Smart Contract Intermediate Representations and Discussion}
Other intermediate representations for smart contracts have been proposed.
Scilla~\cite{scilla} is the intermediate representation used by the Zilliqa blockchain. 
It is based on the notion of communicating automata and contains a translation to the Coq proof assistant to prove safety and liveness properties. 
While Scilla is a promising representation, and it does possess a translation from Solidity code, it is not clear if it would be able to represent the entire Solidity language or only a subset.
Michelson~\cite{michelson} is the intermediate representation used in the Tezos blockchain. 
It is a stack-based representation, making it less suitable for static analysis than SlithIR.
IELE~\cite{iele} is the representation developed for the Cardano blockchain.
IELE is modeled in the K framework, and can take advantage of K existing analyses. 
IELE works at a lower-level than SlithIR and is, therefore, complementary to it.
For example, in IELE, state variables are represented through stores and reads to the storage memory rather than variables, which makes their analysis more complicated, while it allows a more accurate gas estimation.
The Solidity authors are working on an intermediate representation called YUL~\cite{yul}, which is still a work in progress at the time of writing.
While adding an intermediate representation within the Solidity compiler is a good step towards a safer language, YUL does not provide the same level of information as does SlithIR.
For example, YUL does not possess an SSA representation, and is written as a list of nested expressions, making the construction of static analyses less straightforward.

SlithIR possesses some remaining limitations and has room for improvement.
The principal omission is the lack of formal semantics, which would allow more rigorous analyses.
Another limitation is that the representation is too high-level to accurately reflect low-level information, such as the gas computation.

\section{Evaluation And Comparison to State-of-the-Art Tools}\label{sec:evaluation}

We performed an evaluation of three aspects of Slither: (1) vulnerability detection, (2) missing optimization detection, and (3) source-code exploration. 
Slither and the other state-of-the-art tools are evaluated using real-world contracts.
Most of the state-of-the-art tools are focused on the detection of security issues, and therefore, (1) is the core focus of our evaluation.

\subsection{Vulnerability Detection Evaluation}\label{sec:evaluation-vulnerabilities}

We compare Slither (release 0.5.0)  with other open-source static analysis tools to detect vulnerabilities in Ethereum smart contracts: Securify (revision \texttt{37e2984}), SmartCheck (revision \texttt{4d3367a}) and Solhint (release 1.1.10). We decided to focus our evaluation exclusively on tools' reentrancy detectors, since reentrancy is one of the oldest, most well understood, and most dangerous security issues. This detector is  available in all the tools we evaluated. 

We do not consider tools based on dynamic analysis, such as Manticore, Oyente or Mythril, as they are known to have scalability issues and do not compete directly with static analysis tools.

\subsubsection{Experiments} \label{subsec:experiments}

We evaluate the static analysis tools using two experiments: in the first, we used two famous contracts vulnerable to reentrancy, DAO\cite{DAO} and SpankChain\cite{spank},  
to verify if the tools can rediscover the issues actually exploited. 
In the second experiment, we used the $1000$ most used contracts (those with the largest number of transactions), for which Etherscan~\cite{etherscan} provides the source code. 
Several other evaluations in the literature~\cite{oyente,maian,smartcheck} use a direct sample of contracts in the blockchain instead. 
However, we found that such a selection is biased in an unhelpful way, as most of the contracts deployed on Ethereum are essentially tests, lacking the complexity (and code quality) that widely used contracts have.
As a result, most of these contracts are likely to contain trivial-to-detect bugs not reflecting real-world situations, which can skew results.  A previous effort to compare Solidity analysis tools (both static and dynamic) proposed even more selective choice of contracts, based on Zeppelin audits, but this resulted in a set of only 28 contracts to analyze~\cite{dika2017ethereum}; we were able to use a larger set by restricting the scope of comparison, rather than the contracts selected.

\subsubsection{Metrics}

The tools are evaluated on three metrics: performance, robustness, and accuracy. 
The performance metric indicates how fast a tool runs on each contract. 
It should include every step needed to analyze a smart contract. If a tool requires a contract to be compiled, the compilation time is included in this metric. 
The robustness metric indicates how often each tool fails to parse, decompile, or analyze a smart contract.  Finally, the accuracy metric measures both number of contracts flagged and number of false positives detected. 

%We decide to focus the evaluation of this measurement exclusively using the reentracy detectors, since it is one of the oldest, better understood and most dangerous security issues. This detector is also available in all the tools we are evaluating. 

\subsubsection{Reentrancy Revisited} \label{subsec:background-reentrancy}
A function in a smart contract is said to be reentrant if it contains a call to an external contract that can be used to re-enter the function itself. 
A reentrancy can lead to a variety of exploitations, including loss of ether, when a state variable is changed only \emph{after} an external call. %denial-of-service when the contract enters an invalid state.
This issue became famous during the DAO hack~\cite{DAO} and SpankChain\cite{spank}.
%A function allows a reentrant call it can be induced to perform a call to the same function. 
%The intuition behind reentrancy is that it could produce unexpected behavior in a smart contract, since it allows an attacker to 
%A function allows a reentrant call it can be induced to perform a call to the same function. 
%The intuition behind reentrancy is that it could produce unexpected behavior in a smart contract, since it allows an attacker to 
%call a function when the developer did not expect it: when another call was not completed.
%This issue can be use to break some invariants in the contract code and allow a variety of unexpected behaviors, including lost of ether and
%denial-of-service when the contract enters an invalid state.

Figure \ref{fig:reentrancy-1} shows a classic example of reentrancy.
\begin{figure}	
\begin{lstlisting}[basicstyle=\small]
mapping (address => uint) balances;
function withdrawBalance() public {
  if(!(msg.sender.call.value(balances[msg.sender])())) {
    throw;
  }
  balances[msg.sender] = 0;
}
\end{lstlisting}
\caption{Reentrancy example}
\label{fig:reentrancy-1}
\end{figure}
%\gus{explain every line and how it can be exploited}
The exploitability of reentrancy issues is very well known, most vulnerability detection tools for Ethereum smart contracts issue a warning when they detect code like that in Figure \ref{fig:reentrancy-1}.
Reentrancy patterns are not uncommon, but in most of cases, the reentrancy is not severe for two reasons:
\begin{itemize}
\item It requires special privileges: only the owner can act maliciously.
\item It is benign: the "exploit" has the same effect as two successive calls.
\end{itemize}

During our evaluation, if a tool flags a benign reentrancy in a contract, it is considered a false positive.

\subsubsection{Experiment 1}\label{subsec:experiment-1}
%To evaluate the capacity of the tools to detect real reentrancy on complex source code, we evaluate their capacity to detect reentrancies on the DAO and the SpankChain contract. 
%Both source codes are available on the not-so-smart-contract repository\footnote{\url{https://github.com/trailofbits/not-so-smart-contracts}}. 
%We used a modified version of the DAO contract (\jo{link to code}), as the original version does not compile with recent solc version and requires solc 0.3.6, which is not anymore supported by all the tools (including Slither).

The last two columns of Table~\ref{table:stats} show whether the DAO and SpankChain vulnerabilities were found by each tool. 
It is worth noting that several reentrancies are reported by the tools in both examples. However, we consider the result valid only if the tool can detect the actual reentrancy that was used during the exploitation of the contract. Slither is the only tool capable of finding these two real-world cases of reentrancy.

\newcommand{\checked}{$\checkmark$}%
\newcommand{\unchecked}{\ding{55}}%

\begin{table*}[t]
\begin{centering}
\begin{tabular}{|p{0.16\textwidth}|p{0.29\textwidth}|c|c|c|c|}
\cline{3-6} 
\multicolumn{2}{c|}{} & Slither  & Securify  & SmartCheck  & Solhint\tabularnewline
\hline 
\multirow{3}{0.16\textwidth}{\centering{}Accuracy} & False positives  & $10.9\%$  & $25\%$  & $73.6\%$  & $91.3\%$ \tabularnewline
\cline{2-6} 
 & Flagged contracts  & 112  & 8  & 793  & 81 \tabularnewline
\cline{2-6} 
 & Detections per contract  & $3.17$  & $2.12$  & $10.22$  & $2.16$ \tabularnewline
\hline 
\hline 
\multirow{2}{0.16\textwidth}{\centering{}Performance} & Average execution time  & $0.79\pm1$  & $41.4\pm46.3$  & $10.9\pm7.14$  & $0.95\pm0.35$ \tabularnewline
\cline{2-6} 
 & Timed out analyses  & $0\%$  & $20.4\%$  & $4\%$  & $0\%$ \tabularnewline
\hline 
\hline 
\centering{}Robustness & Failed analyses  & $0.1\%$  & $11.2\%$  & $10.22\%$  & $1.2\%$ \tabularnewline
\hline 
\hline 
\multirow{2}{0.16\textwidth}{\centering{}Reentrancy examples} & DAO & \checked  & \unchecked  & \checked  & \unchecked \tabularnewline
\cline{2-6} 
 & Spankchain & \checked  & \unchecked  & \unchecked  & \unchecked \tabularnewline
\hline 
\end{tabular}
\par\end{centering}
\caption{Summary of the evaluation results\label{table:stats}}
\end{table*}

%\subsubsection{Experiment 2: Dataset}\label{subsec:dataset}

%To perform the evaluation of the static analysis tools, we wanted to use a very large dataset of contracts' source codes comprising more than 40 thousands contracts crawled from Etherscan. This dataset was compiled by a Consensys employee and is publicly available~\cite{dataset-big}. The Figure \ref{fig:loc} summarizes the amount of Solidity lines of code of each contract in this dataset. 
%However, using a very large contract could introduce a noticeable bias.  For instance, if the developer deploys a contract to test, realizes it does not work as expected or contains a severe security issue, then the contract is immediately abandoned. Thus, only a very small amount of contracts are killed when they are not necessary or obsolete \gus{citation needed, maybe jay knows about this}. 
%We decided to use a smaller dataset contained only smart contracts actively used, so we selected a subset of the large dataset using by the number of transactions of each contract. We restricted our smaller dataset to first $1000$ contract with the largest number of transactions. \gus{update if we change them} \gus{ this could introduce the bias for older contracts, how to normalize given the age of each contract?} 

\subsubsection{Experiment 2}\label{subsec:experiment-2}

For the second experiment, using a dataset of $1000$ contracts, the tools were run on each contract with a timeout of 120 seconds, using only reentrancy detectors. We manually disabled other detection rules to avoid the introduction of bias in the measurements.  

\begin{itemize}
\item \textbf{Performance:} The \emph{Average execution time} and \emph{Timed-out analyses} rows in Table \ref{table:stats} summarize performance results: Slither is the fastest tool, followed by Solhint, SmartCheck, and Securify. 
In our experiments, Slither was typically as fast as a simple linter. 
%It it important to note that the all information collection regarding the time taken by the tools to run includes the compilation of the contract, if it is needed. 
%In the case of Slither, compiling a contract is absolutely necessary to parse and analyze it. Other tools, such as Solhint and SmartCheck, parse Solidity source code or analyze precompiled contracts, such as Securify. 

\item \textbf{Robustness}:  The \emph{Failed analyses} row in Table \ref{table:stats} summarizes robustness results: Slither is the most robust tool, followed by Solhint, SmartCheck, and Securify. 
Slither failed only for $0.1\%$ of the contracts, while Solhint failed for around $1.2\%$. 
SmartCheck and Securify are less robust, failing $10.22\%$ and $11.20\%$ of the time, respectively. 

\item \textbf{Accuracy}: The \emph{False positives}, \emph{Flagged contracts} and \emph{Detections per contract} rows in Table \ref{table:stats} summarizes accuracy results.  

Due to time constraints, we manually reviewed a random sample of at least 50 of the flagged contracts for each tool, rather than all potentially benign reentrancies.
%Reentrancies that are equivalent to calling a function two times are classified as false positives.

Our experiments show that Slither is the most accurate tool, with the lowest false positive rate: $10.9\%$, followed by Securify with $25\%$. 
SmartCheck and Solhint have high false positive rates of $73.6\%$ and $91.3\%$ respectively. 
Additionally, we include the number of contracts for which at least one reentrancy is detected (\textit{flagged} contracts) and the number of findings on average per \textit{flagged} contract.
%\jo{the detection per contract only include flagged contracts or all the contracts?}. 
On one hand, SmartCheck flags a large number of contracts, confirming its high false positive rate. On the other hand, Securify flags a very small number of contracts, which indicates that the tool fails to find a number of true positives found by other tools.
\end{itemize}

\subsubsection{Discussion}
In these experiments, Slither outperforms the other analyzers for detecting reentrancy by detecting real-world bugs while maintaining a low false positive rate.
However, we should note that our experiments are limited in many ways.
The false positives rate for Securify is based on a very low number of findings (8).
Additionally, several contracts in our dataset share fragments of code, creating a bias in the results. For instance, the false positives from Slither result from only a few distinct functions that are duplicated across multiple contracts.
Despite these limitations, the experiment shows that Slither is more accurate and mature than the other tools.

We also found that SmartCheck and Solhint have a large number of false positives due to a lack of in-depth understanding of Solidity.
For example, the Solidity attribute \textit{super}, which allows calling an inherited function, is taken as an external call by these tools.

\subsection{Optimization Detection Evaluation}\label{sec:evaluation-optimization}

To test the optimization capacity of Slither, we select the detector finding variables that should have been declared as constants.
If a variable is declared constant, it will not take space in the storage of the contract, and will require fewer instructions when used. 
As a result, using constant variables when possible reduces the code size (and thus the cost to deploy the contracts), as well as the usage cost.
To the best of our knowledge, Slither is the only available tool capable of finding this code optimization.

We applied the detector on two datasets: (1) the 1000 contracts selected for experiment 2 (Section~\ref{sec:evaluation-vulnerabilities}), and (2) 35.000 contracts from ~\cite{etherscan}.
We found that 54\% of the contracts from (1) and 56\% from (2) contain one or more variables that could have been declared as constant. Some of these variables are never accessed and could have been entirely removed.

This result shows that Slither can efficiently be used to detect practical code optimization.

\subsection{Code Understanding Comparison}\label{sec:evaluation-code-understanding}

Slither printers provide a variety of visual outputs to improve code understanding.
The closest tool that provides similar features is Surya\footnote{\url{https://github.com/ConsenSys/surya}}. Surya parses the contract AST and offers a set of outputs. It only performs a syntactic analysis and does not provide any semantic analysis.

Table~\ref{table:code-understanding} shows a comparison of Slither printers with Surya features.
\begin{table*}
\begin{centering}
\begin{tabular}{|>{\centering}p{0.3\textwidth}|c|c|}
\hline
                                 & Slither           & Surya        \\ \hline\hline
Contract Summary                 & \textit{contract-summary } & \textit{describe    }   \\ \hline
Functions Summary          & \textit{function-summary } & \textit{            }   \\ \hline
Inheritance dependencies          & \textit{inheritance      } & \textit{dependencies}   \\ \hline
Inheritance graph                & \textit{inheritance-graph} & \textit{inheritance }   \\ \hline
Call graph                       & \textit{call-graph       } & \textit{graph       }   \\ \hline
AST                              & \textit{                 } & \textit{parse       }   \\ \hline
Functions call trace             & \textit{                 } & \textit{ftrace      }   \\ \hline
Authorization and access summary & \textit{vars-and-auth    } & \textit{            }   \\ \hline
Report                           & \textit{human-summary    } & \textit{mdreport    }   \\ \hline
\end{tabular}
\par\end{centering}
\caption{Comparison of Slither printers with Surya}
\label{table:code-understanding}
\end{table*}
Both tools provide similar features, except than Surya outputs a textual representation of the AST and the function call trace, while Slither yields information about variables read and written.
Both tools can generate a report, but they differ in their contents. 
Surya shows contracts and functions name, as well as the mutability and modifiers calls, while Slither shows the number of bugs found, the code complexity (based on cyclomatic complexity), and provides high-level information for specific context based on a set of heuristics (such as the minting limitation in case of ERC20 tokens). 

Slither includes the information provided by Surya, but can integrate more advanced information, as it has an in-depth understanding of the codebase.

\subsection{Threats to Validity}

The most important threat to validity is that our primary experiment is limited to reentrancy detection over a limited set of contracts.  The measure we used to determine a meaningful set of contracts (number of transactions) is reasonable~\cite{measurepop}, but not standardized by the community. It was unclear how to use some very recently proposed frameworks for smart contract analysis, e.g. SmartAnvil~\cite{smartanvil}, which were thus omitted from our experiments.

In order to make our results reproducible, and allow checking of our code and analysis methods, we have made our experimental setup available\footnote{\url{https://gist.github.com/ggrieco-tob/748bca8a0d166e026ea717e225a4fbf9}}.

\section{Conclusions and Future Work}\label{sec:conclusions}

We presented Slither, an open source static analysis framework for smart contracts.
Slither is fast, robust, accurate, and provides rich information about smart contracts.
Our framework leverages SlithIR, an intermediate representation designed for practical static analysis on Solidity code.
Slither is the only platform that can be used at the same time to find bugs, suggest code optimizations, increase code understanding for a given contract.
It was built to be easily extended, and serve as a basis upon which third-party tools can be implemented.

We evaluated the bug finding capacity of Slither by comparing its performance on reentrancy bugs against other available state-of-the-art tools.
We found that Slither outperforms the other tools in terms of performance, robustness, and accuracy. 

There are several directions that we could take to improve Slither.
As the tool is daily used during audits, a first improvement is the integration of new issue detectors.
Optimization could be applied to SlithIR to generate more efficient code.
Symbolic execution or bounded model checking could be built on top of our intermediate representation, allowing easy access to formal verification for bug detectors, and loop-bound based worst-case gas cost analyses.
The Slither analyzer was originally designed to target Solidity code, but both the platform and the intermediate representation can be adapted for another smart contract language, such as Vyper.
Finally, a translation from SlithIR to EVM or Ewasm bytecode would allow Slither to serve as a compiler.

\bibliographystyle{splncs04}
\bibliography{bib/ether}

\end{document}